\documentclass[11pt,nofootinbib,prd,letterpaper]{revtex4}
\usepackage[latin9]{inputenc}
\usepackage{hyperref}
\usepackage{graphicx}
\usepackage{epstopdf}
\usepackage{amsmath}

\makeatletter
\usepackage{slashed}
\makeatother

\newcommand{\mpsi}{m_\psi}
\newcommand{\mchi}{m_\chi}

\def\VEV#1{\langle #1 \rangle}

\begin{document}

\title{Charged-Particle Decay at Finite Temperature}

\author{Andrzej Czarnecki$^1$, Marc Kamionkowski$^{2,3}$, Samuel
K.\ Lee$^{2,3}$, and Kirill Melnikov$^3$}
\affiliation{$^1$Department of Physics, University of Alberta,
Edmonton, AB, Canada T6G 2G7}
\affiliation{$^2$California Institute of Technology, Mail Code 350-17, Pasadena, CA
91125, USA}
\affiliation{$^3$Department of Physics and Astronomy, Johns Hopkins University, Baltimore,
MD 21218, USA}

\begin{abstract}
Radiative corrections to the decay rate of
charged fermions caused by the presence of a thermal bath of
photons are calculated in the limit when temperatures are below
the masses of all charged  particles involved.  The cancellation 
of finite-temperature infrared
divergences in the decay rate is described  in detail.
Temperature-dependent radiative corrections 
to a two-body  decay of a hypothetical charged fermion and 
to electroweak decays of a muon $\mu \to e \nu_\mu \bar \nu_e$ are given.
We touch upon possible implications of these results 
for charged particles in the early
Universe.
\end{abstract}

\maketitle

\section{Introduction}

Precise predictions for decay rates of  charged particles
might be of interest in a variety of cosmological contexts that introduce
long-lived particles with electric charge.  These include scenarios for
modified big-bang nucleosynthesis \cite{Kaplinghat:2006qr,Kohri:2006cn,Kawasaki:2007xb,Kusakabe:2007fu,Jedamzik:2007cp,Takayama:2007du,Jittoh:2011ni} and
small-scale-power suppression \cite{Sigurdson:2003vy,Profumo:2004qt,Kohri:2009mi} and
mechanisms for dark-matter detection wherein a charged
quasistable heavier particle is produced
\cite{Albuquerque:2003mi,Bi:2004ys,Ahlers:2006pf,Albuquerque:2006am,Ahlers:2007js,Reno:2007kz,Ando:2007ds,Canadas:2008ey,Albuquerque:2008zs}.

\begin{figure}[t]
\centering
\parbox[c]{1.5in}{\includegraphics[width=1.5in]{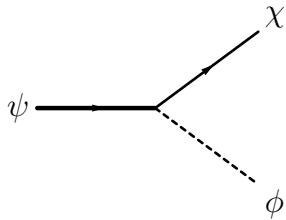}}
\caption[The diagram for the decay $\psi \to \chi\phi$]{The diagram for the decay $\psi \to \phi \chi$. 
Note that we shall distinguish between the $\psi$ and $\chi$ particles in the diagrams below by representing them with thick and thin solid lines, respectively.}
\label{fig:tree}
\end{figure}

\begin{figure}[t]
\centering
$\begin{array}{c@{\hspace*{20mm}}c}
\parbox[c]{1.5in}{\includegraphics[width=1.5in]{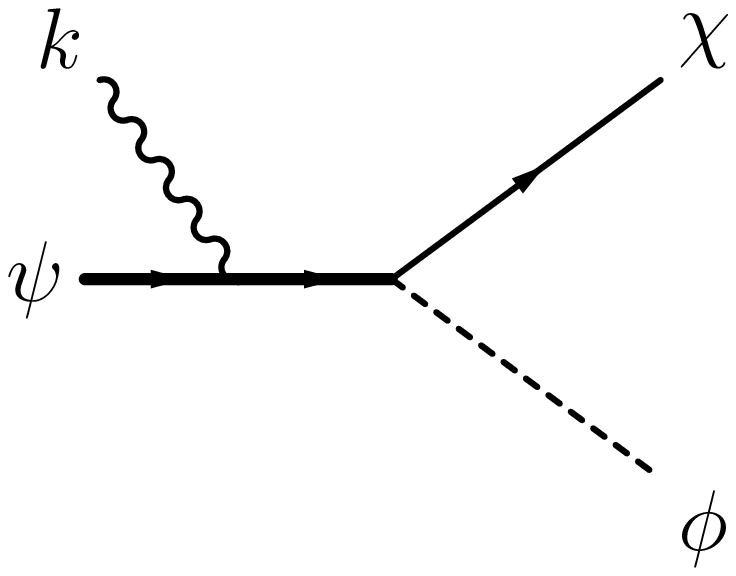}} &
\parbox[c]{1.5in}{\includegraphics[width=1.5in]{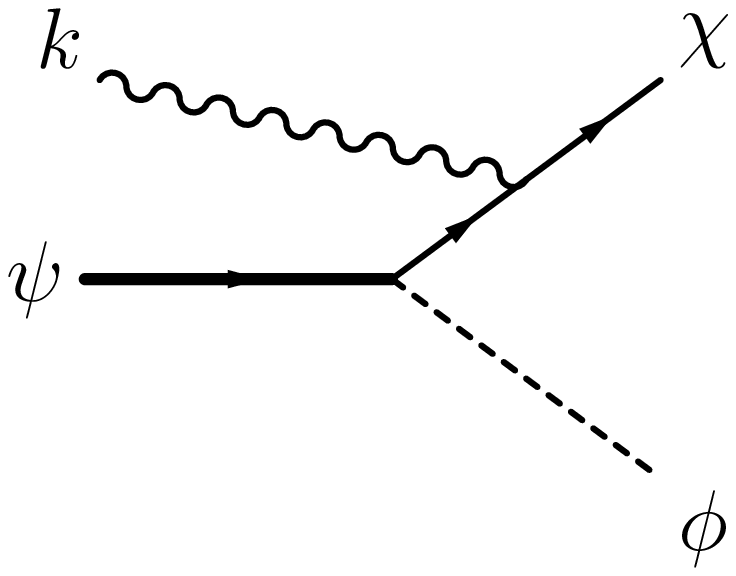}} \\
{\rm s-channel} & {\rm u-channel}
\end{array}$
\caption[The two diagrams via which absorption of a photon can lead to induced
 $\psi$ decay (or $\chi$ and $\phi$ production)]{The two diagrams via which absorption of a photon can lead to induced
 $\psi$ decay (or $\chi$ and $\phi$ production).}
\label{fig:abs}
\end{figure}

In the early Universe, decays of charged particles occur in a thermal bath 
whose very presence seems to affect decay rates in a peculiar way. 
Indeed, consider the decay of a hypothetical charged particle $\psi$ to two
lighter particles $\chi$ and $\phi$. The leading-order 
Feynman diagram is shown in  Figure~\ref{fig:tree}.  
In a thermal bath of photons, the process $\gamma\psi
\rightarrow \chi\phi$  also occurs and modifies the vacuum decay rate.
Hence, the vacuum 
decay rate of a particle $\psi$  must be augmented
in the cosmological context by the inclusion of the rate for the
``induced'' decay wherein the unstable particle absorbs a
thermal photon (see Figure~\ref{fig:abs}).
It is easy to see that a naive computation of the 
diagrams in Figure~\ref{fig:abs} leads to a divergent result. This  divergence 
is of the infrared type -- it appears 
when the energy of the absorbed photon becomes very small.   In this paper 
we discuss in detail how, when all possible processes that modify 
the vacuum decay rate are taken into account, the infrared divergences cancel out. This 
is a finite-temperature analog of the celebrated cancellation of infrared  
divergences in QED pointed out by Bloch and Nordsieck long ago
\cite{Bloch:1937pw}.
  
Although thermal effects have been computed for static
thermodynamic quantities such as the effective potential, the
free energy, the pressure, and so on \cite{bellac}, sometimes to
very high orders in the perturbative expansion in QCD and QED
(for recent  reviews, see Refs.\
\cite{Kraemmer:2003gd,Moller:2010xw}), less is known about
finite-temperature corrections to cross sections and decay
rates.  Radiative corrections to dynamical scattering and decay
processes at a finite temperature $T$ are peculiar for three
reasons. First, as pointed out already, 
if $T\ne0$, new processes 
involving absorption and emission of particles from the heat bath
contribute to cross-sections and decay rates. 
The second complication is that the preferred reference
frame defined by the heat bath spoils Lorentz
invariance. Third, thermal averages and loop integrals over
Bose-Einstein distributions introduce infrared divergences that
are powerlike, rather than logarithmic.

Pioneering studies of radiative corrections to neutron
$\beta$-decays at finite temperature were first described in
Refs.~\cite{Dicus:1982bz,Cambier:1982pc,Baier:1988xv,Baier:1989ub},
in the context of big-bang nucleosynthesis.  In
Ref.~\cite{Donoghue:1983qx}, the finite-temperature decay rate
of a  neutral Higgs boson into two charged leptons was first
computed.  These and subsequent papers
\cite{Donoghue:1984zz,Kobes:1984vb,Johansson:1985if,Ahmed:1986hy,Masood:1987mi,Grandou:1988fv,Keil:1988sp,Keil:1988rv,Grandou:1990ir,Altherr:1991cq,Weldon:1991eg,Niegawa:1991qk,Niegawa:1993fu,LeBellac:1995rs,Indumathi:1996ec}
illustrated the cancellation of infrared divergences and
clarified many important features of radiative corrections.
They also discussed the issue of radiative corrections 
 that are enhanced by  the logarithms of small masses of 
final-state charged particles. Such terms are known to cancel 
in total decay rates at zero temperature \cite{Kinoshita:1962ur}, 
but the situation at  finite temperature is less clear.

Most of the papers just described 
dealt only with a neutral initial state; in such a case, the problem 
of an infinite decay rate induced by absorption of very soft photons 
by the initial state does not 
occur.  Here we discuss the calculation of radiation 
corrections for a charged initial state, where this issue can not be avoided. 
For simplicity, we begin by considering  a toy model of
charged-fermion decay and focus on the low-temperature case.  We
introduce the toy model in Section~\ref{sec:real} and calculate the
decay rate induced by real-radiation scattering processes in
the thermal bath (i.e., absorption and emission of photons),
showing how infrared divergences arise.  In
Section~\ref{sec:virtual}, we compute  the virtual radiative
corrections to the decay rate.  In Section~\ref{sec:total}, we
sum the real and virtual corrections to find the total
finite-temperature decay rate, and demonstrate the cancellation
of the divergences at first order in perturbation theory.  We
carry out an analogous analysis for muon decay in
Section~\ref{sec:muon}.  
We conclude and
consider the implications for charged particles in the early
Universe in Section~\ref{sec:conclusions5}.

\section{Toy model} \label{sec:real}

We shall first discuss a simple model to illustrate the nature
of the infrared divergences and their cancellation.  Consider
the process  $\psi \to \chi\phi$, the decay of a heavy charged
fermion $\psi$ to a light charged fermion $\chi$ and a massless
neutral scalar $\phi$ via the interaction
\begin{equation}
{\cal L} \supset g \phi \bar{\psi} L \chi + \textrm{h.c.}\,,
\end{equation}
depicted in Figure~\ref{fig:tree}.  Here $L=(1-\ensuremath{\gamma_{5}})/2$.  We shall assume that the charge of both fermions is the elementary charge $e = \sqrt{4\pi\alpha}$, and that the mass ratio $\epsilon \equiv \mchi/\mpsi$ 
is small.

At $T=0$, the tree-level amplitude for the decay $ \psi \to \chi \phi$ 
is given by
\begin{equation} \label{eq:tramp}
{\cal M}_{\rm tr} = g \bar{u}_\chi L u_\psi\,.
\end{equation}
This amplitude gives the ${\cal O}(g^2\alpha^0)$ zero-temperature 
 decay rate,
\begin{equation}
\widetilde{\Gamma}_0 = \Gamma_0 (1-\epsilon^4)\,, \label{eq:fulltree}
\end{equation}
where we have defined
$\displaystyle
\Gamma_0 \equiv \frac{g^2}{32 \pi} \mpsi.
$
We will state our subsequent results for the temperature-dependent 
decay rate  in terms of $\Gamma_0$.

On account of radiative corrections and finite-temperature effects, the decay rate  
can be written as a 
triple series in $\tau = T/m_\psi$, $\epsilon = m_\chi/m_\psi$, 
and the fine-structure constant  $ \alpha$. Unless explicitly stated otherwise, 
we work in the low-temperature approximation $\tau \ll \epsilon \ll 1$. Our goal is to compute 
relative corrections to the decay rate that scale as $\alpha \tau^2$. In those 
terms, we shall set $\epsilon \to 0$. 
Note that terms   of the form  $\tau/\epsilon$ do not   
appear in the total decay rate.

\subsection{Photon absorption}

We now consider the process $\gamma\psi \to \chi\phi$, the induced decay of $\psi$ 
in a thermal bath of photons. This process can
occur via the two diagrams shown in Figure~\ref{fig:abs}.  For a photon with
4-momentum $k=(\omega, {\bf k})$, where $\omega = |{\bf k}|$,
the tree-level amplitudes for these channels are
\begin{eqnarray} \label{eq:absamps}
{\cal M}_{\rm abs,s} &=& \frac{e g}{s -\mpsi^2} \bar{u}_\chi L (\slashed{p}_\psi + \slashed{k} + \mpsi) \slashed{\epsilon}_\gamma u_\psi\,,\\ 
{\cal M}_{\rm abs,u} &=& \frac{e g}{u -\mchi^2} \bar{u}_\chi \slashed{\epsilon}_\gamma (\slashed{p}_\chi - \slashed{k} + \mchi) L u_\psi,
\end{eqnarray}
giving the total amplitude ${\cal M}_{{\rm abs}} = {\cal M}_{\rm
abs,s}+{\cal M}_{\rm abs,u}$.
We use the amplitude to compute the cross-section for $\gamma\psi \to 
\chi\phi$ by the standard procedure. We
assume that the $\psi$ particle is at
rest with respect to the photon bath and express 
 the result in terms of the energy of the photon 
$w \equiv \omega / \mpsi$. \footnote{We are implicitly considering a cosmological scenario
 in which the heavy $\psi$ particle has decoupled
from the thermal bath and is out of equilibrium.  Furthermore, we assume that the massless
$\phi$ particle is also not thermalized in the bath (which may be the
case if the coupling $g$ is very weak and $\psi$ decay is the dominant mode of $\phi$ production, for example).  Similar assumptions will also be made in the
case of muon decay discussed in Section~\ref{sec:muon}.} We find 
\begin{eqnarray} \label{eq:absxs}
     \sigma_{{\rm abs}}(w) &\equiv& \frac{1}{2\mpsi} \frac{1}{2
     |\omega|} \int\! {\rm dLIPS}_{\chi\phi}\, (2\pi)^4
     \delta^4(p_\psi+k-p_\chi-p_\phi) \VEV{|{\cal M}_{{\rm
     abs}}(k)|^2} \nonumber\\
         &=& \Gamma_0 \frac{\alpha \pi}{\mpsi^3 w^3}  \rho(w)\,,
\end{eqnarray}
where
\begin{equation} \label{eq:rho}
     \rho(w) \equiv \left(1+2w+2w^2\right) \ln
     \frac{1+2w}{\epsilon^2} - \left(2+4w+3w^2\right),
\end{equation}
at leading order in $\epsilon$.  Note that $\rho \propto w^0$
and $\sigma_{{\rm abs}} \propto w^{-3}$ as $w \to 0$.

To compute corrections to the decay rate of a particle $\psi$ due to 
the absorption of thermal photons from the heat bath, we need to integrate
the cross-section $\sigma_{{\rm abs}}(w)$ in Eq.~(\ref{eq:absxs}) multipled with the 
average occupation number for thermal photons.
We find 
\begin{eqnarray}
\label{eq:abs2}
 \Gamma^T_{{\rm abs}} &=& \int {\rm d}w\, \frac{dn_\gamma}{dw}(w) \sigma_{\rm abs}(w) \nonumber\\
&=& g_\gamma \int \frac{d^3{\bf k}}{(2\pi)^3} f_B(\omega) \sigma_{{\rm abs}}(w) \nonumber\\
&=& \frac{\alpha}{\pi} \Gamma_0 \int_0^\infty\! \frac{dw}{w} f_B(\omega) \rho(w)\,,
\end{eqnarray}
where $f_B(\omega) = 1/(e^{\omega/T}-1)= 1/(e^{w/\tau}-1)$ is the 
Bose-Einstein distribution function 
and $g_\gamma=2$ is the number of independent photon polarizations. 
Since $f_B \propto w^{-1}$ and $\rho \propto w^0$ as $w
\to 0$, we see that the integrand in Eq.~(\ref{eq:abs2}) is
proportional to $w^{-2}$ as $w \to 0$, and hence $\Gamma^T_{{\rm
abs}}$ indeed has a powerlike infrared divergence, implying an
infinite decay rate.

As in the case of infrared divergences at zero temperature, the infinite rate is  unphysical.  We arrived at
this unphysical result because we considered only the photon
absorption process $\gamma\psi \to \chi\phi$ in the calculation
of the finite-temperature $\psi$ decay rate.  However, as
we shall demonstrate, we cannot consider this absorption process
independently of other processes that also result in $\psi$
decay.  In particular, we must also take into account the
finite-temperature rates of the radiative decay
process $\psi \to \gamma\chi\phi$ and the decay process $\psi
\to \chi\phi$.  At finite temperature, the emission of photons
in the first process is stimulated by the presence of photons in
the thermal bath.  At the same order in $\alpha$, the second
process is affected by $T$-dependent 
additions to the
virtual-photon propagator.
When all of these processes are included in the calculation, all
$T$-dependent infrared divergences cancel to yield a finite
rate.  The nature of this cancellation is 
similar to zero-temperature cancellations of
infrared divergences in QED, as described in a classic paper  by 
Bloch and Nordsieck \cite{Bloch:1937pw}.
We shall now demonstrate this cancellation to
${\cal O}(g^2 \alpha)$.

\begin{figure}[htbp]
\centering
$\begin{array}{c@{\hspace*{20mm}}c}
\parbox[c]{1.5in}{\includegraphics[width=1.5in]{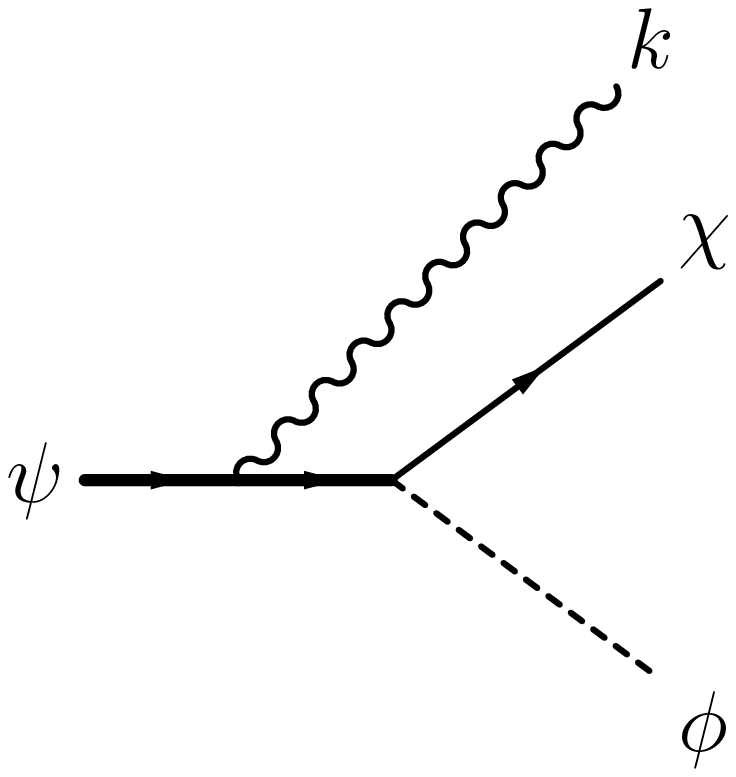}} &
\parbox[c]{1.5in}{\includegraphics[width=1.5in]{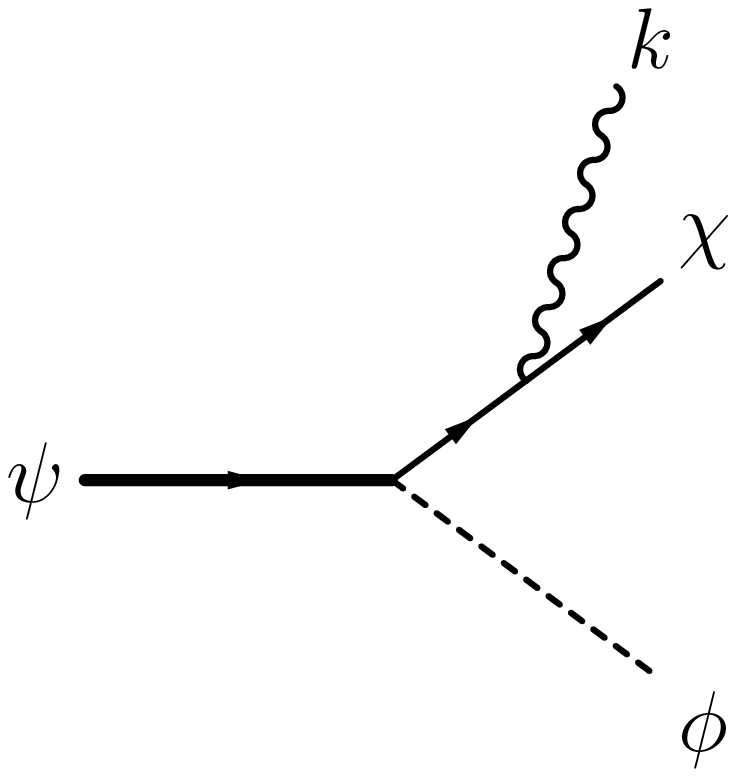}}
\end{array}$
\caption[The diagrams for radiative $\psi$-decays]{The diagrams for radiative $\psi$-decays.
Note that the topologies of the diagrams are identical to those
for absorption in Figure~\ref{fig:abs}, with the exception of the
photon line placement.  This results in the relation given in
Eq.~(\protect\ref{eq:absemeq}).}
\label{fig:em}
\end{figure}

\subsection{Photon emission}

We begin by considering  
the photon-emitting radiative 
decay process $\psi \to \gamma\chi\phi$.  The two contributing diagrams 
are shown  in Figure~\ref{fig:em}. 
Comparing these diagrams to those in Figure~\ref{fig:abs}, we see that the 
amplitudes for emission are formally equivalent to those for 
absorption given in 
Eq.~(\ref{eq:absamps}) if we make the substitutions
$
k \leftrightarrow -k$ and
$\epsilon_\gamma \leftrightarrow \epsilon^\ast_\gamma\,$.
This crossing symmetry gives the photon-emission amplitude 
from the photon-absorption amplitude,
\begin{equation} \label{eq:absemeq}
\VEV{|{\cal M}_{{\rm em}}(k)|^2}=g_\gamma \VEV{|{\cal M}_{{\rm abs}}(-k)|^2}\,,
\end{equation}
where the factor of $g_\gamma$ arises because 
we do not average over photon polarizations  on the left-hand side.  
The ${\cal O}(g^2\alpha)$ $T$-dependent part of the rate for this process is 
then given by
\begin{equation} \label{eq:em}
\Gamma^T_{{\rm em}} = \frac{1}{2\mpsi} \int\! {\rm dLIPS}_{\gamma\chi\phi}\, f_B(\omega) (2\pi)^4 \delta^4(p_\psi-k-p_\chi-p_\phi) \VEV{|{\cal M}_{{\rm em}}(k)|^2}\,,
\end{equation}
where the factor of $f_B(\omega)$ comes from the $T$-dependent part of the $(1+f_B)$ Bose-enhancement factor for the final-state photons.
Comparing  Eq.~(\ref{eq:em}) to
Eq.~(\ref{eq:abs2}), and using
Eqs.~(\ref{eq:absxs})  and (\ref{eq:absemeq}), it can
be shown that the expression for the emission rate is very
similar to that for the absorption rate given in Eq.~(\ref{eq:abs2}),
\begin{equation}
\Gamma^T_{{\rm em}} = \frac{\alpha}{\pi} \Gamma_0 \int_0^{1/2}\! 
\frac{dw}{w} f_B(\omega) \rho(-w)\,. \label{eq:emone}
\end{equation}
We see that the only differences are $\rho(w) \to \rho(-w)$, arising from the $k \to -k$ substitution used to switch the absorbed photon to an emitted photon, and the limits of integration.  The upper limit reflects that the emitted photon is limited by the kinematics to have $w < 1/2$ in the final state, whereas an absorbed photon is allowed to have any energy in the initial state.  Note that this is further reflected in the fact that $\rho(w)$ is defined only 
for $-1/2 < w < \infty$.

\subsection{Real-radiation corrections}

We are now in position to calculate the total rate of $\psi$ decay due to processes involving either the absorption or emission of real photons,
\begin{equation} \label{eq:real}
\Gamma^T_{{\rm real}} = \Gamma^T_{{\rm abs}}+\Gamma^T_{{\rm em}} = 
\frac{\alpha}{\pi} \Gamma_0 \int_0^{\infty}\! 
\frac{dw}{w} f_B(\omega) \rho_{{\rm real}}(w)\,,
\end{equation}
where $\rho_{{\rm real}}(w) = \rho(w) + \theta(1/2-w) \rho(-w)$.
Exact integration over $w$ in the above formula is complicated because of the Bose-Einstein factor.  However, if we consider the low-temperature case $\tau \ll 1$, then $f_B(\omega)$ is only non-negligible for $w \lesssim \tau \ll 1/2$.  We can thus use the approximation 
$\theta(1/2-w) \to 1$ and integrate 
Eq.~(\ref{eq:real}) by expanding $\rho_{{\rm real}}(w)$ in a Taylor series in $w$.  
This approximation picks up all the terms that are suppressed by powers of $\tau$, but it misses the exponentially suppressed terms of ${\cal O}(e^{-1/\tau})$.
The ${\cal O}(g^2\alpha)$ result is then
\begin{equation} \label{eq:realresult}
\Gamma^T_{{\rm real}} = \frac{\alpha}{\pi} \Gamma_0 \left[\left(-4-4\ln\epsilon + {\cal O}(\epsilon^2 \ln \epsilon)\right) J_{-1} + \left(-2-8\ln\epsilon + {\cal O}(\epsilon^2)\right) J_1 \tau^2 + {\cal O}(\tau^4,e^{-1/\tau})\right]\,,
\end{equation}
where we have defined the integrals
\begin{equation}
J_n \equiv \lim_{x_0 \to 0}\int_{x_0}^\infty\! dx\, x^n f_B(x T) \theta(x-x_0)\,.
\end{equation}
Note that for $n > 0$, $J_n = {\rm Li}_{n+1}(1)\Gamma(n+1)$ is
finite\footnote{A few terms read 
 $\displaystyle J_1 = \frac{\pi^2}{6}$, $J_3 = \frac{\pi^4}{15}$,
$J_5 = \frac{8\pi^6}{63}$, $J_7 = \frac{8\pi^8}{15}$.}.
The infrared-divergent part of the decay rate due to real-radiation processes is thus given by the $J_{-1}$ term in
Eq.~(\ref{eq:realresult}); we shall now proceed to show that it
is canceled by corresponding terms in the virtual corrections to
the decay rate.

\begin{figure}[t]
\centering
$\begin{array}{cc}
\parbox[c]{1.5in}{\includegraphics[width=1.5in]{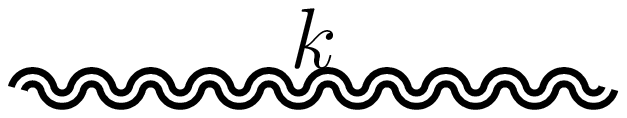}} & = -2\pi g_{\mu\nu} f_B\!\left(|k^{0}|\right)\delta\!\left(k^{2}\right)
\end{array}$
\caption[The $T$-dependent part of the photon propagator]{The $T$-dependent part of the photon propagator.}
\label{fig:photon}
\end{figure}

\section{Virtual corrections} \label{sec:virtual}

As previously mentioned, the rate of the decay process $\psi \to \chi\phi$ is affected by $T$-dependent additions to the virtual  corrections that enter at ${\cal O}(g^2\alpha)$.  These affect both the vertex correction shown in Figure~\ref{fig:vert} and the charged-fermion self-energies shown in Figure~\ref{fig:self}.  At finite temperature, the bare propagator for the virtual photons that appear in these diagrams is modified compared to the zero-temperature case,
 \begin{equation}
-\frac{ig_{\mu\nu}}{k^{2}+i0}\to-g_{\mu\nu}\left[\frac{i}{k^{2}+i0}+2\pi f_B\left(|k^0|\right)\delta\left(k^{2}\right)\right]\,. \label{eq:photonprop}
\end{equation}
The first term in this equation is the usual $T=0$ photon propagator; its effects 
are accounted for in  conventional zero-temperature perturbation theory. The 
second term in the right-hand side of Eq.~(\ref{eq:photonprop}) 
leads to temperature-dependent corrections to the decay rate (see Figure~\ref{fig:photon}). 
We note that the temperature-dependent contribution to the photon propagator 
accounts for interactions  of real, on-shell photons from the thermal bath with the 
charged fermions in the initial and final states.  In particular, 
it represents processes in which a photon from the thermal bath is absorbed by either of the fermions, while simultaneously another photon is emitted by either of the fermions with the exact same momentum and polarization as the initial photon.\footnote{We can see how the term arises by expanding the photon field in terms of creation and annihilation operators in the usual manner.  At $T=0$, $aa^{\dagger}$ terms generate the usual propagator, but the $a^{\dagger}a$ terms proportional to the particle number vanish.  At finite temperature, the latter terms are instead proportional to the occupation number $f_B$ and hence are nonvanishing.}

\subsection{Vertex correction}

\begin{figure}[t]
\centering
$\begin{array}{c}
\parbox[c]{1.5in}{\includegraphics[width=1.5in]{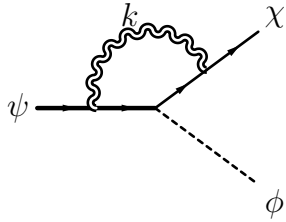}}
\end{array}$
\caption[The diagram contributing to the $T$-dependent part of the ${\cal O}(\alpha)$ correction to the vertex]{The diagram contributing to the $T$-dependent part of the ${\cal O}(\alpha)$ correction to the vertex.}
\label{fig:vert}
\end{figure}

We first consider the $T$-dependent part of the ${\cal O}(\alpha)$ correction to the vertex, 
shown in Figure~\ref{fig:vert}.  The $T$-dependent part of the relevant amplitude is given by
\begin{equation}
{\cal M}^T_{\rm vert} = -e^2 g \int\! \frac{{\rm d}^4k}{(2\pi)^3} F(k^0,{\bf k}) f_B(|k^0|)\delta(k^2)\,,
\end{equation}
where
\begin{equation}
F(k^0,{\bf k}) \equiv \frac{\bar{u}_\chi \gamma^\mu (\slashed{p}_\chi - \slashed{k} + \mchi) L (\slashed{p}_\psi - \slashed{k} + \mpsi) \gamma_\mu u_\psi}{\left[(p_\chi - k)^2 -\mchi^2\right] \left[(p_\psi - k)^2 -\mpsi^2\right]}\,. \label{eq:Fk}
\end{equation}
We may use the properties of the gamma matrices to simplify this expression.  Integration over $k^0$ removes $\delta\left(k^{2}\right)$,
\begin{equation}
{\cal M}_{\rm vert}^{T}=-e^{2}g \int\!\frac{{\rm d}^{3}{\bf k}}{(2\pi)^{3}2\omega}\left[F(\omega,{\bf k})+F(-\omega,{\bf k})\right] f_B(\omega)\,,
\label{powerdiv}
\end{equation}
where $\omega=|{\bf k}|$.  Now, by changing the variable of
integration from ${\bf k}\to-{\bf k}$ in the second term in the
sum enclosed in brackets, the sum becomes $[F(k) + F(-k)]$.
Examining Eq.~(\ref{eq:Fk}), we see that the numerator of this
sum is then independent of $k$, since $k^2=0$ and terms that are
linear in $k$ cancel.  However, the denominator of the sum
retains its quadratic dependence on $k$, since it is
proportional to $(p_\chi\cdot k)(p_\psi\cdot k)$.  By counting
powers of $\omega$ and recalling that $f_B \propto \omega^{-1}$
at small values of $\omega$, we observe that the  integral in Eq.~(\ref{powerdiv})
indeed has a powerlike divergence.

We can then take the interference of this amplitude with the tree-level amplitude given in Eq.~(\ref{eq:tramp}).  The $T$-dependent part of the vertex correction to the decay rate then follows by taking the usual spin-sum average and evaluating the resulting integral over ${\bf k}$,
using the appropriate kinematics of the 2-body decay in the rest frame of the $\psi$ particle.  Writing the result in terms of the integrals $J_n$ as before, we find at ${\cal O}(g^2\alpha)$
\begin{equation}
\Gamma^T_{\rm vert} = \frac{\alpha}{\pi} \Gamma_0 \left[\left(4\ln \epsilon + {\cal O}(\epsilon^2 \ln \epsilon)\right)J_{-1}   \right]\,. \label{eq:vertresult}
\end{equation}
Comparing to the decay rate due to real-radiation processes in
Eq.~(\ref{eq:realresult}), we see that part of the divergent
$J_{-1}$ term is indeed canceled.

\subsection{Self-energy corrections} \label{sec:self}

\begin{figure}[t]
\centering
\parbox[c]{1.5in}{\includegraphics[width=1.5in]{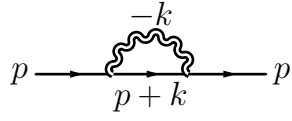}}
\vspace*{-12mm}
\caption[The diagram contributing to the $T$-dependent part of the fermion self-energy]{The diagram contributing to the $T$-dependent part of the fermion self-energy.}
\label{fig:self}
\end{figure}

The remaining part  of the infrared divergence is canceled by the $T$-dependent
corrections to the fermion self-energy $\Sigma^T(p)$, which
enter via the virtual photon propagator in Figure~\ref{fig:self} and lead to
$T$-dependent contributions to the full (dressed) fermion
propagator $S_F^T(p)$.  At $T=0$, these self-energy
contributions are conveniently treated through the  mass shift $\delta m$,
and the  wave-function
renormalization factor $Z_2$. The wave function renormalization factors  
are usually obtained as derivatives of the self-energy
$\Sigma$ with respect to $p$, evaluated on the mass shell
$p^{2}=m^{2}$.  This treatment relies on the fact that at $T=0$,
$\Sigma$ depends only on the momentum of the particle
$p$. Unfortunately, this feature is violated at finite
temperature because the thermal bath introduces a preferred
reference frame. As a result, the self-energy of a particle at
rest and the self-energy of a particle in motion are not related
in an obvious way.

We will need expressions for the self-energy of both the $\psi$ and $\chi$ fermions.  We can consider more generally the $T$-dependent part of the self-energy $\Sigma^{T}(p)$ of a fermion with electric charge $e$ and $T=0$ physical mass $m$.  This calculation has been discussed extensively in the literature; 
below, we loosely follow 
the formalism laid out in Ref.~\cite{Altherr:1989fc}.  
Our ultimate goal will be to use the expression for $\Sigma^{T}$ to show that, 
in the limit $p^2 \to m_T^2$,  the full finite-temperature fermion propagator takes the form
\begin{equation}
S_F^T(p) = Z_2^T \frac{i \sum_s u_s^T(p) \bar{u}_s^T(p)}{p^2 - m_T^2}\,. \label{eq:propform}
\end{equation}
That is, the pole of the propagator is shifted to the finite-temperature physical mass $m_T$, and the fermion wave functions are given by $\Psi_s(p) = \sqrt{Z_2^T/2p^0} u_s^T(p) e^{-ip\cdot x}$, where $Z_2^T$ and $u_s^T(p)$  are the finite-temperature wave-function renormalization factor and the finite-temperature spinor, respectively.\footnote{Note that we assume that renormalization has already been carried out at $T=0$.}  This form implies that self-energy corrections to the decay rate will follow from three distinct sources: (1) matrix elements will be multiplied by a factor of $Z_2^T$ for each external fermion line, (2) the shift in the physical mass, which will effectively modify the fermion phase-space, and (3) the spinor completeness relation will be modified, affecting the evaluation of spin sums.

To this end, we start by finding the self-energy $\Sigma^T$ at ${\cal O}(\alpha)$.  
We take $p$ to be the off-shell fermion momentum and $-k$ to be the momentum of the photon in the loop, as shown in Figure~\ref{fig:self}.  
The self-energy reads 
\begin{equation}
\Sigma^T(p) = 2 e^2 \int\!\frac{{\rm d}^4 k}{(2\pi)^3} \frac{\slashed{p}+\slashed{k}-2m}{(p+k)^2-m^2} f_B(|k^0|)\delta(k^2)\,. \label{eq:selfenergy}
\end{equation}
It is convenient to decompose $\Sigma^T(p)$
as 
\begin{equation}
\Sigma^T(p) = \slashed{p}c_B(p) - 2m c_B(p) + \slashed{K}(p)\,, \label{eq:selfenergydecomp}
\end{equation}
where
\begin{eqnarray}
\label{eq:cB}
c_B(p) &\equiv& 2 e^2 \int\!\frac{{\rm d}^4 k}{(2\pi)^3} \frac{f_B(|k^0|)\delta(k^2)}{(p+k)^2-m^2}\,, \\
K^\mu(p) &\equiv& 2 e^2 \int\!\frac{{\rm d}^4 k}{(2\pi)^3}\; k^\mu \frac{f_B(|k^0|)\delta(k^2)}{(p+k)^2-m^2} \nonumber\\
&\underset{p^2=m^2}{\to}& \frac{\alpha}{\pi}J_1 \frac{T^2}{|\bf{p}|} \left(L_p, 
\frac{\bf{p}}{|\bf{p}|}\left[\frac{p^0}{|\bf{p}|} 
L_p - 2\right]\right)\,, 
\end{eqnarray}
and $\displaystyle L_p = \ln\frac{p^0 + |\bf{p}|}{p^0 - |\bf{p}|}$.
We can now use these results for $\Sigma^T$ to find the full (dressed) finite-temperature fermion propagator $S_F^T(p)$ at ${\cal O}(\alpha)$ in the usual way,
\begin{eqnarray}
S_F^T(p) &=& \frac{i}{\slashed{p} - m - \Sigma^T} \nonumber\\ 
	     &=& \frac{i[\slashed{p}(1-c_B) + m(1-2c_B)-\slashed{K}]}{p^2(1-2c_B) - m^2 (1-4c_B) - 2p\cdot K + {\cal O}(\alpha^2)}\,. \label{eq:fullprop2}
\end{eqnarray}
Examining the denominator of Eq.~(\ref{eq:fullprop2}), we see
that it can be simplified by defining
\begin{eqnarray}
\delta m_T^2 &\equiv&  2 e^2 \int\!\frac{{\rm d}^4 k}{(2\pi)^3} f_B(|k^0|)\delta(k^2) \nonumber\\ 
	&=& \frac{2\pi}{3} \alpha T^2. \label{eq:deltamsqeval}
\end{eqnarray}
Also,  since $-2p \cdot k = -[(p+k)^2-m^2] + (p^2 - m^2) + k^2$, we can write
\begin{equation}
-2p\cdot K = -\delta m_T^2 + (p^2-m^2) c_B\,. \label{eq:pdotK}
\end{equation}
Finally, 
we can expand in $(p^2-m^2)$ around the on-shell momentum $\widehat{p}$ (which satisfies $\widehat{p}^2=m^2$) by writing
\begin{equation}
c_B(p) = \widehat{c_B} + (p^2-m^2) \widehat{c_B}' + {\cal O}\left((p^2-m^2)^2\right)\,, \label{eq:cBexp}
\end{equation}
where
$ \widehat{c_B} \equiv c_B(\widehat{p}) = 0 $ and 
\begin{eqnarray}
\widehat{c_B}' &\equiv&  \frac{dc_B}{dp^2}(\widehat{p}) \nonumber\\ 
 &=& -2 e^2 \int\!\frac{{\rm d}^4 k}{(2\pi)^3} \frac{f_B(|k^0|)\delta(k^2)}{((\widehat{p}+k)^2-m^2)^2} \left(1 +  \frac{d(2p \cdot k)}{dp^2}(\widehat{p})\right) \nonumber\\ 
&=& -\frac{\alpha}{\pi} \frac{J_{-1}}{m^2}\,. \label{eq:cBhatpeval}
\end{eqnarray}
We note that the vanishing of the $\widehat{c_B}$ coefficient follows from 
the antisymmetry of the integrand for $c_B$ in Eq.~(\ref{eq:cB}) at 
$p = \hat p$  w.r.t.  $k \to -k$; for the same reason, the
derivative term in the integrand in Eq.~(\ref{eq:cBhatpeval})  vanishes as
well.

We are now in position to recover the form of the propagator
advertised in Eq.~(\ref{eq:propform}), by using
Eqs.~(\ref{eq:cBexp}) and (\ref{eq:pdotK}) in
Eq.~(\ref{eq:fullprop2}), and keeping only 
${\cal O}(\alpha)$ terms there.  We find the result
\begin{equation}
S_F^T(p) = (1-2m^2 \widehat{c_B}') \frac{i(\slashed{p} + m - \slashed{K})}{p^2 - m^2 - \delta m_T^2}\,.
\end{equation}
Comparing this expression with Eq.~(\ref{eq:propform}) and using 
Eqs.~(\ref{eq:cBhatpeval}) and (\ref{eq:deltamsqeval}), we obtain 
\begin{eqnarray}
Z_2^T &=& 1-2m^2 \widehat{c_B}' \nonumber\\ &=&  1 + 2 \frac{\alpha}{\pi} J_{-1}\,, \label{eq:Z2eval}\\
m_T^2 &=& m^2 + \delta m_T^2 \nonumber\\ &=& m^2 + \frac{2\pi}{3}\alpha T^2\,, \label{eq:massshift}\\
\sum_s u^T_s(p) \bar{u}^T_s(p) &=& \slashed{p} + m - \slashed{K}(p)\,, \label{eq:spinsum}
\end{eqnarray}
where in the last expression the momentum-dependent results of
Eq.~(\ref{eq:cB}) are to be used in evaluating spin sums.
These finite-temperature relations affect the decay rate in the
three aforementioned ways; we shall now calculate each of their
contributions separately. 

First, the finite-temperature wave-function renormalization
factor $Z_2^T$ simply affects the tree-level decay rate
$\widetilde{\Gamma}_0$ of Eq.~(\ref{eq:fulltree}) as an overall
multiplicative factor; one factor enters for each external
fermion line.  This yields the ${\cal O}(g^2\alpha)$
temperature-dependent contribution
\begin{equation}
\Gamma^T_{Z_2} =  \frac{\alpha}{\pi}\Gamma_0\left[ \left(4 + {\cal O}(\epsilon^4)\right) J_{-1}\right]\,. \label{eq:Z2result}
\end{equation}
Combining this with Eqs.~(\ref{eq:realresult}) and
(\ref{eq:vertresult}), we see that this contribution cancels the
remaining infrared-divergent $J_{-1}$ part of the total decay
rate.

Second, Eq.~(\ref{eq:massshift}) 
results in a shift of the pole of the fermion propagator to $p^2=m_T^2$.  
Since the pole masses of the fermions define the leading-order rate, the mass shifts 
${\Delta m_i \equiv m_{T,i} - m_i \approx \delta m_{T,i}^2 / 2 m_i}$ lead to the following ${\cal O}(g^2\alpha)$  ``phase-space'' correction 
\begin{eqnarray}
\Gamma_{\rm{ph}}^T &=& \frac{\partial \widetilde{\Gamma}_0}{\partial \mpsi} \Delta \mpsi + \frac{\partial \widetilde{\Gamma}_0}{\partial \mchi} \Delta \mchi \nonumber\\ 
		&=& \frac{\alpha}{\pi} \Gamma_0 \left[\left(2 + {\cal O}(\epsilon^2)\right) J_1 \tau^2  \right]\,. \label{eq:phresult}
\end{eqnarray}

Finally, the finite-temperature spin-sum relation found in
Eq.~(\ref{eq:spinsum}) modifies the calculation of matrix
elements.  Note from Eq.~(\ref{eq:cB}) that the relation is
actually momentum-dependent, and must be computed for both the
$\psi$ particle (at rest) and the $\chi$ particle (with energy
$p^0 = \mpsi (1+\epsilon^2)/2$).  The leading-order contribution
to the decay rate is then found by using the finite-temperature
spin-sum relation in the tree-level calculation, giving the
${\cal O}(g^2\alpha)$ temperature-dependent part,
\begin{equation}
     \Gamma_K^T = \frac{\alpha}{\pi} \Gamma_0 \left[\left(-2 + 8
     \ln \epsilon + {\cal O}(\epsilon^4)\right) J_1
     \tau^2\right]\,. \label{eq:Kresult}
\end{equation}

The total ${\cal O}(g^2\alpha)$ self-energy correction is then
given by the sum of these three effects \[Eqs.~(\ref{eq:Z2result})--(\ref{eq:Kresult})\],
\begin{eqnarray}
\Gamma_\Sigma^T &=& \Gamma_{Z_2}^T + \Gamma_{\rm ph}^T + \Gamma_K^T \nonumber\\ 
		&=& \frac{\alpha}{\pi} \Gamma_0 \left[\left(4+{\cal O}(\epsilon^4)\right)J_{-1} + \left(8 \ln \epsilon+ {\cal O}(\epsilon^2)\right) J_1 \tau^2\right]\,. \label{eq:selfresult}
\end{eqnarray}

\section{Total decay rate in the toy model}
\label{sec:total}

We are now in position to present the final formula for the 
decay rate of the hypothetical fermion $\psi$ in a thermal bath. We consider the  
low-temperature limit $T \ll \mpsi, m_\chi$ and include
contributions from processes involving both real photons and
virtual photons; the latter category includes the vertex
correction and corrections arising from the self-energy of
charged fermions.  The total decay rate is the sum
of these contributions given in 
Eqs.~(\ref{eq:realresult},\ref{eq:vertresult},\ref{eq:selfresult}).  
We remind the reader that 
our calculation is performed in the 
approximation $ \tau \ll \epsilon \ll 1$ and 
that  we are interested in the leading 
${\cal O}(\alpha \tau^2)$ 
temperature-dependent 
correction to the rate. 
We find
\begin{eqnarray}  \label{eqn:totalresult}
\Gamma_{\rm tot}^T &=& \Gamma_{\rm real}^T + \Gamma_{\rm vert}^T + \Gamma_{\Sigma}^T \nonumber\\ &=&
- \alpha  \frac{\pi}{3}\tau^2 \Gamma_0  + 
{\cal O}(\tau^2 \epsilon^4,\tau^4,e^{-1/\tau})\,.
\end{eqnarray}
We see that  all the infrared-divergent  terms proportional to the integrals $J_{-1}$ 
cancel out in the total rate. We note that  this statement remains  
valid  if exact $\epsilon$-dependence  of the rate is restored. 
Furthermore, we find in Eq.~(\ref{eqn:totalresult}) that all  the terms that contains logarithms of the mass 
ratio $\ln \epsilon$ cancel in the correction to the total rate, in contrast 
to  individual contributions in 
Eqs.(\ref{eq:realresult},\ref{eq:Kresult}). 
Cancellation of such terms in the zero-temperature case 
follows from the Kinoshita-Lee-Nauenberg theorem 
\cite{Kinoshita:1962ur,Lee:1964is}.  We are not 
aware of a general proof of a similar cancellation 
at  a {\it finite} temperature, 
so it is important to watch for such terms in explicit $T \ne 0$
computations.


\section{Muon decay $\mu
\to e \nu_\mu \bar{\nu}_e $ } \label{sec:muon}

In this Section, we present the temperature-dependent correction 
to the muon decay rate  at low temperature. 
The details of the calculation are similar to the preceding discussion 
of the toy model. The main difference is that the muon decay 
is a three-body process, so that integration over the 
phase-space of final-state particles is more complex.

The muon decay to electron and neutrinos  is described by an effective  
Lagrangian 
\begin{equation}
     {\cal L}\supset
     \frac{4G_{F}}{\sqrt{2}} \; \bar{e}\gamma^{\rho}L 
     \nu_e\; \bar{\nu}_{\mu}\gamma_{\rho}L\mu,
\end{equation}
where $G_{F}$ is the Fermi constant.  The leading-order, zero-temperature 
decay rate reads 
\begin{equation}
\widetilde{\Gamma}_0 = \Gamma_0  \left(1 - 8\epsilon^2 - 24 \epsilon^4 \ln \epsilon  + 8\epsilon^6 -\epsilon^8 \right)\,, \label{eq:treemuon}
\end{equation}
where now
$\displaystyle
\Gamma_{0}\equiv \frac{G_{F}^2 m_{\mu}^{5}}{192\pi^{3}}
$ and $\epsilon \equiv m_e/m_\mu$. 
Similar to the toy-model case, the radiative corrections to the rate are 
given by the sum of real photon emission/absorption 
corrections, the vertex corrections 
and the self-energy corrections. For future reference, we show those 
corrections separately.

The contribution to the decay rate from real photon emission/absorption reads 
\begin{eqnarray}
\Gamma_{\rm real}^T = \frac{\alpha}{\pi} \Gamma_{0}\left[\left(-\frac{17}{3}-4\ln\epsilon\right) J_{-1} + \left(-\frac{70}{3}-32\ln\epsilon\right) J_{1}\tau^{2}
 \right]\,. \label{eq:realresultmuon}
\end{eqnarray}
Note that here and below we keep 
only the leading term in $\epsilon$ for each power of  $\tau \equiv T/m_\mu$, and consistently 
neglect all powers of $\tau$ beyond $\tau^2$. The result for the 
vertex correction reads 
\begin{equation}
\Gamma_{{\rm vert}}^T=\frac{\alpha}{\pi}\Gamma_{0} \left[\left(\frac{5}{3}+4\ln\epsilon\right)J_{-1}\right]\,. \label{eq:vertresultmuon}
\end{equation}
We note that the  temperature dependence 
in Eq.~(\ref{eq:vertresultmuon}) 
is exact and  that higher-order 
terms in $\tau$ do not appear there. 

The self-energy correction to the fermion propagator was discussed 
in the previous section and much of that discussion remains valid. 
For this reason, we just summarize the result.
The total self-energy correction is given by
\begin{equation}
\Gamma_\Sigma^T = \frac{\alpha}{\pi} \Gamma_0 \left[ 4 J_{-1} + \left(\frac{64}{3} + 32 \ln \epsilon\right) J_1 \tau^2\right]\,. \label{eq:selfresultmuon}
\end{equation}

The final result for  temperature-dependent radiative corrections to the muon 
decay is given by the sum of the three contributions of
Eqs.(\ref{eq:realresultmuon},\ref{eq:vertresultmuon},\ref{eq:selfresultmuon}).
 Including also the
$T=0$ radiative corrections
\cite{Behrends:1955mb,Berman:1958ti}, we find
the final ${\cal O}(\alpha,\tau^2,\epsilon^0)$ result 
\begin{equation}
\Gamma_{\mu \to e \nu \bar \nu } = \Gamma_0 \left\{
     1+\frac{\alpha}{\pi}\left[\left(\frac{25}{8} -
     \frac{\pi^{2}}{2}\right) - \frac{\pi^{2}}{3} \tau^2 \right]
     \right\}.\label{eq51}
\end{equation}
We note that ${\cal O}(\alpha \tau^2)$ correction to the rate 
for the muon decay is identical to the analogous correction 
to the two-body fermion decay rate in the toy model, suggesting the possibility 
of deriving and understanding this result in a simpler fashion.
We also note that our  result Eq.~(\ref{eq51}) disagrees with the one given in
Ref.~\cite{Jagannathan:1990hb}.  

Part of the discrepancy can be traced to the issue of mass singularities. 
Indeed, in Ref.~\cite{Jagannathan:1990hb}, the $\ln \epsilon$ terms 
are present even at ${\cal O}(\alpha \tau^2)$ contributions to the rate 
but, as follows from our analysis, such terms cancel when 
all contributions are taken into account.
Nevertheless, as pointed out already, whether mass singularities cancel in the rate if  
higher-order terms in $\tau$ are accounted 
for is an open question.  When we extend the calculation of the muon decay rate to include  
${\cal O}(\alpha \tau^4)$ terms, we find that Eq.~(\ref{eq51}) is modified 
by 
\begin{equation}
\Delta \Gamma_{\mu \to e \nu \bar \nu }  = - \frac{\alpha}{\pi} 
\Gamma_0 \frac{64 \pi^4 \tau^4}{45} \left ( 2 \ln \epsilon + \frac{1}{3} 
\right ),
\label{eq52}
\end{equation}
which shows the logarithmic sensitivity to the electron-to-muon
mass ratio.
In the low-temperature 
regime that we consider in this paper $\tau \ll \epsilon \ll 1$, 
there exist  more important corrections to Eq.~(\ref{eq51}) than the 
ones displayed in Eq.~(\ref{eq52}). However, most of the ``more relevant'' 
corrections involve powers of the mass ratio $\epsilon$, while Eq.~(\ref{eq52}) 
shows logarithmic sensitivity to $\epsilon$, a unique feature in the low-temperature regime. It is interesting to point out that by 
relaxing the relationship between 
the temperature $T$ and the mass of the charged particle in the final state 
(the electron), we obtain new sources of mass logarithms related  
to the thermal Fermi-Dirac distribution of {\it fermions} in the heat bath. 
Complete analysis of the corrections to the muon decay rate for the 
{\it intermediate-temperature regime} $ m_e \ll T \ll m_\mu$ -- where proper 
interplay of bosonic and 
fermionic temperature-dependent corrections becomes  important -- 
is beyond the scope of the present  paper. 

\section{Conclusions}
\label{sec:conclusions5}

Long-lived charged particles appear in a variety of scenarios
for early-Universe physics and dark matter.  In all cases, these
long-lived particles are bathed for a long time in a gas of
photons, giving rise to the possibility of induced decays through
processes such as those shown in Figure~\ref{fig:abs}.  If the
rate of this induced process is large, then the cosmological 
effects of these long-lived charged particles may be substantially modified.

A naive evaluation of the rate for these induced decays leads to 
a result which diverges as the frequency of the photon 
in the heat  bath that induces the decay vanishes.  As with infrared 
divergences at zero temperature, a proper calculation of  the decay 
rate 
requires accounting for all degenerate processes. Once this is done, 
the infrared divergences cancel, leading to small correction to the decay 
rate. By considering a simple  toy model with a two-body final 
state and a realistic process -- muon decay -- with a three-body 
final state, we found a universal 
leading finite-temperature correction 
$\displaystyle \delta \Gamma/\Gamma_0  
= -\frac{\alpha}{\pi} \frac{\pi^2 T^2}{3m^2}$, 
where  $m$ is the mass of the decaying particle. 

In this paper, we focused on discussing infrared divergences 
in decays of charged particles in the thermal bath. This issue 
can be sharply defined by considering temperatures that are  
small compared to masses of decaying particles and their decay 
products.   An interesting set of questions arises  if we depart 
from the low-temperature regime and consider 
the ``intermediate''-temperature scenario, where the mass of the decaying 
particle is large and masses of decay products are small, compared to the 
heat-bath temperature.  In such a case, radiative corrections 
enhanced by the logarithms of the mass ratios can become numerically 
important in the context of a variety
of scenarios that occur in the early Universe.  For example,
light gravitinos that arise in theories of supergravity with
gauge-mediated supersymmetry breaking may be produced by the decay of
short-lived charged staus to taus \cite{Dine:1981za,Dimopoulos:1981au,Nappi:1982hm,AlvarezGaume:1981wy,Dine:1994vc,Dine:1995ag}. 
Temperatures greater than the tau
mass will then fall into the  ``intermediate''-temperature scenario. 
If mass-enhanced corrections  are indeed present, 
large modifications of  the stau decay
rate and the production rate of light gravitino dark matter
become conceivable. Such modifications may  
affect the early-Universe thermal history and have implications 
for collider phenomenology \cite{Feng:2010ij}.

Furthermore, in some regions of the supersymmetric parameter
space, the process of co-annihilation is important in the
determination of the dark-matter relic abundance after
freeze-out \cite{Binetruy:1983jf,Griest:1990kh}.  Freeze-out occurs roughly at
temperatures $T \sim m_{\rm SUSY} / 20$ that may be greater than
the masses of some of the products of supersymmetric particle
decay, so this indeed presents another intermediate-temperature scenario.
Finite-temperature effects and mass singularities may then
become important in determining the individual scattering and
decay rates for charged particles, which would be important if
one is interested in the detailed thermal history of these
particles.  However, note that the final dark-matter relic
abundance is most likely not strongly affected by
finite-temperature effects, since only the co-annihilation rates
(and not other scattering or decay rates) enter the calculation
\cite{Edsjo:1997bg}.  For example, in
Ref.~\cite{Wizansky:2006fm} it was found that finite-temperature
corrections to co-annihilation occur only at the $10^{-4}$
level. Finite-temperature effects may also affect neutrino decoupling
\cite{Fornengo:1997wa}.  It has also been suggested that the
original calculations of temperature-dependent corrections 
to  neutron decay are incomplete
\cite{Brown:2000cp}.  Clearly, there remains much work to be
done concerning finite-temperature effects in the early
Universe.

\begin{acknowledgments}
This research was supported at Alberta by Science and Engineering Research Canada (NSERC), at JHU by the NSF
grant PHY-0855365 and JHU startup funds, and at Caltech by DoE
DE-FG03-92-ER40701, NASA NNX10AD04G and the Gordon and Betty
Moore Foundation.
\end{acknowledgments}


\providecommand{\href}[2]{#2}\begingroup\raggedright\endgroup

\end{document}